\documentclass[journal=jctcce,manuscript=article]{achemso}

\usepackage[version=3]{mhchem} 
\usepackage{subcaption}
\usepackage{booktabs}

\usepackage[dvipsnames]{xcolor}
\usepackage{soul}

\author{Nicol\'o Antonini}
\affiliation{Dipartimento di Farmacia, Universit\`a G. d'Annunzio Chieti-Pescara, via dei Vestini, 66100 Chieti, Italy}
\alsoaffiliation{Dipartimento di Chimica, Biologia e Biotecnologie, Universit\`a degli Studi di Perugia, via Elce di Sotto 8, 06123 Perugia, Italy}

\author{Enrico Ronca}
\affiliation{Dipartimento di Chimica, Biologia e Biotecnologie, Universit\`a degli Studi di Perugia, via Elce di Sotto 8, 06123 Perugia, Italy}

\author{Loriano Storchi}
\email{loriano@storchi.org}
\affiliation{Dipartimento di Farmacia, Universit\`a G. d'Annunzio Chieti-Pescara, via dei Vestini, 66100 Chieti, Italy}
\alsoaffiliation{
CNR Institute of Chemical Science and Technologies ``Giulio Natta'' (CNR-SCITEC), 
via Elce di Sotto 8, 06123 Perugia, Italy}

\author{Leonardo Belpassi}
\email{leonardo.belpassi@cnr.it}
\affiliation{
CNR Institute of Chemical Science and Technologies ``Giulio Natta'' (CNR-SCITEC), 
via Elce di Sotto 8, 06123 Perugia, Italy}

\title[An \textsf{achemso} demo]
{Automatic generation of density fitting auxiliary basis sets for all electron Dirac-Kohn-Sham calculations.}

\abbreviations{RQC, ABS, 4c-DKS}

\begin{document}

\begin{abstract}
In this study, we present a general workflow that enables the automatic generation of auxiliary density basis sets for all elements of the 
periodic table (from H to Og) 
to facilitate the general applicability of
relativistic Dirac-Kohn-Sham calculations. It is an important tool for the accurate description of relativistic effects, including spin-orbit coupling, 
in molecules containing heavy elements. The latter are very important in various fields ranging from catalysis to quantum technologies.
The automatic generation algorithm is based on an even-tempered scheme inspired by a previous work by P. Calamicini et al. J. Chem. Phys. 2007, 126, 7077, in which the auxiliary basis sets were generated
 for non-relativistic DFT calculations within the GGA approximation. Here, the algorithm uses basic information from the principal relativistic spinor basis set (exponents and angular momentum values) and includes a simple strategy to account for the high angular momentum of electrons in heavy and superheavy elements.
The workflow developed here allows us to perform extensive automated tests aimed at verifying the accuracy of the auxiliary basis sets in a large molecular data set of about 300 molecules representing all groups and periods of the periodic table.
The results show that our auxiliary basis sets achieve high accuracy, with errors in the Coulomb energies of a few $\mu$-hartree, which are of the same order of magnitude as in the non-relativistic density fitting.
The automatic workflow developed here is general and will be applied in the future for optimization of auxiliary basis sets to include of exact exchange to relativistic approaches. The latter will be a crucial step for the accurate description of spectroscopic properties and the spin-dynamics in molecular systems containing heavy elements.
\end{abstract}

\section{Introduction}

Heavy elements are of great interest in various fields, including catalysis, optoelectronic, spintronics and other quantum technologies.
Accurate simulations of molecular systems containing heavy elements require the correct inclusion of relativistic effects (spin-orbit coupling and scalar relativistic corrections), electron correlation and, for very accurate predictions of spectroscopic constants, also QED corrections\cite{Trond:2022,Matyus:2024}. In particular, spin-orbit coupling may influence laser-induced spin dynamics and unexpected phenomena such as chirality-induced spin selectivity\cite{ciss_chemrev:2024}.
 
The most accurate way to introduce relativity in the modelling of molecular systems is to use the four-component (4c) formalism derived from the Dirac equation. The full 4c formalism is particularly interesting since it
represents the most rigorous way of explicitly treating all the interactions involving spin, which are today of great
technological importance. Furthermore, rigorous relativistic theories offer the natural framework to describe the interaction of particles with electromagnetic fields. As a matter of fact, when determining the electronic structure of molecules containing heavy elements, the effects of electron correlation are just as important as those of the theory of relativity.
These effects are generally not additive and should be treated on the same footing. Due to the large number of electrons that have to be correlated in systems containing heavy atoms, explicit wave function-based electron correlation methods rapidly exceed practicability limits.
A promising approach to include the electron
correlation is given by the relativistic four-component generalization of the Kohn-Sham scheme referred to as the Dirac-Kohn-Sham (4c-DKS) method. 
The practical application of the 4c-DKS methods was hampered by the high computational effort required to properly evaluate the relativistic electron repulsion integrals and all associated properties. To alleviate these limitations, the implementation of density fitting approach was a tremendous breakthrough in reducing the computational burden. The latter was applied in the 4c framework since 2006 in the 4c molecular code BERTHA \cite{BelpB16_20}.
The idea of incorporating the density expansion obtained from the  Coulomb fitting directly into the exchange–correlation functional (in local and semilocal approximation) was also particularly effective in further reducing the cost of DKS calculations\cite{Belpassi_PhysRevB}.
The introduction of the so-called variational density fitting scheme  has proven to be particularly effective in combination with the implementation of various parallelisation and memory distribution schemes
\cite{storchi2010efficient, storchi2013efficient,rampino2014full, belpassi2020bertha}. Recently, we have shown that the DKS method based of density fitting can be easily ported to GPUs.\cite{BERTHA_GPU}

The density fitting scheme is now successfully applied in other 4c implementations\cite{respect:2020}\cite{bagel:2013}. The density fitting was also used for Dirac-Fock using the Coulomb, Gaunt, and full Breit interactions\cite{bagel:2013}.
Despite the proven advantages of density fitting, its systematic application in a 4c-relativistic context is still somewhat limited. The density fitting requires a pre-optimised atomically centred auxiliary basis set.  So far,\cite{BelpB16_20} we have used an atom-optimised auxiliary basis set that follows a very simple two-step optimisation procedure where 1) the exponents  optimised on the spherical atom (see Quiney et al.\cite{Quiney2002}) and 2) angular momentum determined to reduce the error in the Coulomb energy below a required threshold.
In  Ref.s~\cite{respect:2020, bagel:2013}, for example, the auxiliary basis sets used were specifically generated by an adjusted even-tempered algorithm.
In all cases, there was no systematic analysis of the accuracy when the auxiliary basis sets are used in molecules.
Typically, only a few molecular systems are used, leaving a significant gap between the benefits of the density fitting technique and its general applicability in routine calculations. Furthermore, while there are many well-established molecular data sets to test the performance of various auxiliary fitting basis sets in non-relativistic context,  they are much more limited for all electron calculations of molecules containing heavy and superheavy elements.

In recent years, there has been increasing interest in the topic of automatically generated Auxiliary Basis Sets (ABS). One of the advantages of these is that they retain the information of the principal basis set and therefore have greater flexibility compared to optimised ABS, which are typically generated using an energetic minimization criterion for the atoms.
 This is particularly important for applications such as spectroscopic simulations, where the optimised nature of the principal basis set must be retained in the auxiliary set.
Among others, we cite the remarkable works by Lehtola\cite{Lehtola1, Lehtola2}, Stoychev\cite{Stoychev1}, Aquilante et al.\cite{Aquilante1,Aquilante2,Aquilante3}, Hellman et al.\cite{AutoABSTDDFT} and Yang et al.\cite{YANGetal} that help to understand the actual 
state of development in this field.
Most automated generation strategies rely on decomposition techniques, such as Cholesky decomposition, which allow a minimal but sufficient amount of auxiliary functions to be extracted. Alternatively, one can generate auxiliary functions using some direct information from the principal atomic orbital basis set. The implementation is relatively simple, but one may introduce some uncontrolled errors.
In the Cholesky decomposition-based approach, the auxiliary basis set is evaluated on-the-fly, the accuracy is controlled by a single parameter, but an efficient implementation can be non-trivial and the auxiliary basis sets are usually larger than in density fitting. Despite the widespread application in non-relativistic contexts or even in relativistic two-component \cite{Cheng:2024}, the generalisation to 4c is not straightforward. The difficulty arises from the fact that it is necessary to deal with basis functions that have different properties, including those associated with the large and small components of the spinorial structure of the principal basis set.
Recently, Li et al.\cite{RelRIChol} proposed a Cholesky integral decomposition based on the Pauli quaternion representation for the relativistic Dirac-Coulomb Hamiltonian and presented an approximated strategy to evaluate the Cholesky vectors for the (SS|LL) block integrals.
Unfortunately, the error in the evaluation of these integrals is not bounded by the Cholesky threshold used for the large block. However, no variational collapse is observed in actual numerical tests, suggesting that the Cholesky decomposition may offer promising advantages in terms of memory storage and efficiency.

The accuracy of the density fitting approach, and in particular the automatically generated ABS, can only be assessed on the basis of extensive benchmark calculations on a large number of molecules with atoms in different molecular environments and oxidation states.
These benchmarks can be very time consuming. In general, the molecular data sets are not specific to relativistic calculations and heavy and superheavy elements are under-represented.
A notable exception is the molecular data set proposed by Pollak et al.\cite{PolWeigSET} which includes molecules with atoms up to Rn.

In this study, we have developed a fully automated method for generating relativistic auxiliary basis sets, tested the accuracy and analysed the results using a large molecular set of about 290 molecules representing all periods and groups of the periodic table (from H to Og).
While the details of the algorithm are specific to the structure of the auxiliary basis set (Hermite-Gaussian type functions) used in the DKS module of the BERTHA package \cite{pyberthagitrefrel, de2020pyberthart, belpassi2020bertha}, the automated workflow is completely general and can be easily adapted to other quantum chemistry codes. In particular, we have automated the computational processes at each stage of the workflow, i.e. from the initial selection of basis sets to the final validation of the results. The automation has also extended to data manipulation and analysis, where we have used Python-based tools for benchmarking and visualisation.

The paper is organised as follows. Section 2 gives a brief overview of the main theoretical aspects of the DKS method as implemented in BERTHA, with particular reference to the variational density fitting implementation and to the definition of the auxiliary fitting basis set. Section 3 describes the automatic generation algorithm starting from a given principal G-spinor basis set. In Section 4 we describe the fully automated workflow. In Section 5, instead, we present the overall performance in terms of accuracy, measured as mean error (and standard deviation) on the Coulomb energy, using as reference the DKS calculation without density fitting. Finally, we draw some conclusions and perspectives in Section 6.

\section{Theoretical background}

In the present section, we will review the basic theory of the 4-components Dirac-Kohn-Sham formalism, with a focus on the aspects specifically related to the G-spinor basis sets and the variational density fitting approach as implemented in the BERTHA code.
The 4c-DKS equation can be written as
\begin{equation}\label{4cDKS}
\left\{c\mbox{\boldmath $\alpha \cdot p$ \unboldmath} +\beta c^2 
+v^{L}({\bf r})\right\}\Psi_{i}({\bf r}) = E_i\Psi_i({\bf r})
\end{equation}
where $c$ is the speed of light in vacuum, $\boldsymbol{p}$ is the electron
momentum,
\begin{equation}
{\boldsymbol{\alpha}} = \left(
\begin{array}{cc}
0 & {\boldsymbol{\sigma}} \\
{\boldsymbol{\sigma}} & 0
\end{array}
\right)\ \mbox{and}\
{\beta} =
\left(
\begin{array}{cc}
I & 0 \\
0 & -I
\end{array}
\right)
\end{equation}
where ${\boldsymbol{\sigma}}=(\sigma_x,\sigma_y,\sigma_z)$, with $\sigma_q$ is a
$2\times 2$ Pauli spin matrix and $I$
is a $2\times 2$ identity matrix.

The effective longitudinal potential $v^L({\bf r})$ is given by the sum of the scalar external potential generated by the nuclei $v_{ext}[({\bf{r}})]$, the longitudinal Coulomb interaction $v_{H}^{L}[\rho({\bf{r}})]$ and the 
exchange-correlation potential $v_{xc}^{L}[\rho({\bf r})]$. The latter
is  associated to the
longitudinal exchange–correlation energy. Its exact
form is, like that of the corresponding non-relativistic quantity,
unknown and has to be approximated. Typically, non-relativistic exchange-correlation functionals are employed.
Note that the full Breit term, which contributes to the transverse part of the Hartree
interaction, is not considered here \cite{quiney02_5550,belpassi11_12368}. Note that for the large class of time-reversal invariant systems (that is, closed-shell molecules) it does not contribute anyway. 
In BERTHA, the  spinor solution can be expressed as linear combination of G-spinor basis functions ($M^{(T)}_\mu(\textbf{r})$, T=L,S) in the form of
\begin{equation}\label{spinorexpansion}
\Psi_i({\bf{r}}) = \begin{bmatrix}
\sum_{\mu}^{N} c_{i\mu}^L M^{(L)}_\mu(\textbf{r})\\
i\sum_{\mu}^{N} c_{i\mu}^S M^{(S)}_\mu(\textbf{r})
\end{bmatrix}
\end{equation}
G-spinor are  two-component objects derived from spherical Gaussian-type functions.
The small component ($M^{(S)}_\mu(\textbf{r})$) of a G-spinor is defined from the large component ($M^{(L)}_\mu(\textbf{r})$) via the restricted kinetic balance relation. Details for the G-spinor definitions can be found in Refs.\citenum{grant00_022508,grant07,belpassi11_12368}. 
$\mu$  is a collective index that comprises the fine-structure quantum number $k$, the quantum number associated to the $z$ component of the angular momentum, $m_j$, the Gaussian exponent, and the origin of the local coordinate system. It is often convenient to label the G-spinors using the nominal orbital angular momentum label, $l$, because this makes the correspondence with non-relativistic theory more immediate. This is very useful because it allows to build relativistic basis sets starting from a conventional non-relativistic definition of the basis functions. Among their characteristics,  G-spinor basis set does not suffer from the variational problems as the kinetic balance prescription avoids the variational collapse\cite{kineticbalance}. 

In the G-spinor representation of Eq.\eqref{spinorexpansion}, the total charge density can be defined as follows
\begin{eqnarray}\label{totchargerho}
\rho({\bf r}) & = &\sum_{T=L,S}\sum_{\mu \nu}
D_{\mu\nu}^{TT} \rho_{\mu\nu}^{TT}({\bf{r}}) 
\end{eqnarray}
with $T=L, S$. The matrices ${\bf D}^{LL}$ and ${\bf D}^{SS}$ are the large- and small-component density matrices
\begin{equation}
D^{TT}_{\mu \nu}=\sum_a c^{T \ast}_{a \mu}c^{T}_{a \nu}
\end{equation}
where $\{c^{T}_{a\nu}\}$ is the set of coefficients of the molecular spinor expansion (see Eq.\ref{spinorexpansion}), and the sum over $a$ includes only occupied positive energy states (electrons).

Here, we can appreciate a particular aspect of the 4c-DKS scheme implemented in BERTHA. Indeed, because of the specific definition of G-spinor basis we can express the overlap components $\rho_{\mu\nu}^{TT}({\bf r})=M_{\mu}^{(T)\dagger}({\bf r})M_{\nu}^{(T)}({\bf r})$ as the linear combination of Hermite Gaussian Type Functions (HGTF), according to the following equation:
\begin{eqnarray}\label{ovapchardens}
    \rho_{\mu \nu}^{TT}({\bf r}) & = & M_\mu^{(T)\dagger}(\textbf{r}) M_\nu^{(T)}(\textbf{r}) \\
    & = & \sum_{ijk}E^{TT}_{0}[\mu,\nu;i,j,k]H[\alpha,r_A;i,j,k;{\bf r}]  
\end{eqnarray}
 where
\begin{equation}\label{HGTFgeneraleq}
    H[\alpha, r_A; i, j, k; \textbf{r}] = \frac{\partial^i }{\partial x^i} \frac{\partial^j }{\partial y^j} \frac{\partial^k }{\partial z^k} e^{-\alpha |\textbf{r} - \textbf{r}_A|^2}.
\end{equation}
The definition and construction of the $E^{TT}_{0}$  coefficients
enables the efficient analytic evaluation of all multi-centre
G-spinor Coulomb integrals.
We  recall that the full 4c relativistic code BERTHA is built around an efficient algorithm for the analytical evaluation of relativistic electron repulsion integral, 
originally developed by Quiney and Grant more than twenty years ago, ~\cite{belpassi11_12368, quiney97_829, grant00_022508} which represents the relativistic generalization of the  McMurchie-Davidson algorithm \cite{MCMURCHIE1978218,quiney97_829}.
All the 4-components structure of this formalism is retained into the these $E^{TT}_{0}$ coefficients.  Other significant advantages of this approach will emerge  in the course of the following discussion in relation to the density fitting implementation.

The matrix representation of the DKS operator in the G-spinor basis  is 
\begin{equation} \label{DKSmatrix}
{\bf H}_{DKS}=\left[ \begin{array}{cc} {\bf v}^{LL}+{\bf J}^{LL}+{\bf K}^{LL}+ mc^2 {\bf S}^{LL} &  c{\Pi}^{LS} \\ 
             c{\Pi}^{SL}  & {\bf v}^{SS}+{\bf J}^{SS}+{\bf K}^{SS}- mc^2 {\bf S}^{SS} 
\end{array}\right]
\end{equation}

The  generalized eigenvalue matrix to be solved is given by
\begin{equation}
{\bf H}_{DKS} \left[ \begin{array}{c} {\bf c}^L \\ {\bf c}^S \end{array}\right] 
= E \left[ \begin{array}{cc} {\bf S}^{LL} & 0  \\ 0  & {\bf S}^{SS} \end{array}\right]
\left[ \begin{array}{c} {\bf c}^L \\ {\bf c}^S \end{array}\right] 
\end{equation}
where ${\bf c}^T$ are vectors, and ${\bf v}^{TT}$, ${\bf J}^{TT}$
,${\bf K}^{TT}$, ${\bf S}^{TT}$,${\bf \Pi}^{T\bar T}$ are all
matrices.
The matrix elements mentioned in the above equation are defined as follows
\begin{eqnarray}
S_{\mu,\nu}^{TT} &=& 
                   \int M_{\mu}^{(T) \ \dagger}({\bf r}) \
                   M_{\nu}^{(T)}({\bf r}) d{\bf r} \\
\Pi_{\mu,\nu}^{T\bar{T}} & = & \int 
                  M_{\mu}^{(T) \ \dagger}({\bf r})
                  \  \left( \sigma {\bf\cdot p}\right)  \
                  M_{\nu}^{\bar{(T)}}({\bf r}) d{\bf r}  \\
\mbox{ v}_{\mu,\nu}^{TT} &=& \int 
                  M_{\mu}^{(T) \ \dagger}({\bf r}) \ v_{ext}({\bf r}) \ M_{\nu}^{(T)}({\bf r}) d{\bf r}
\end{eqnarray}
\begin{eqnarray}\label{jelement}
J_{\mu,\nu}^{TT} &=& \int 
                  M_{\mu}^{(T) \ \dagger}({\bf r}) \ v_{H}^{L}[\rho({\bf r})]\
                  M_{\nu}^{(T)}({\bf r}) d{\bf r}
\end{eqnarray}
\begin{eqnarray}\label{kelement}
K_{\mu,\nu}^{TT} &=& \int 
                  M_{\mu}^{(T) \ \dagger}({\bf r}) \ 
                  v_{xc}^{L}[\rho({\bf r})] \
                  M_{\nu}^{(T)}({\bf r}) d{\bf r} 
\end{eqnarray}
where $S_{\mu,\nu}^{TT}$, $\Pi_{\mu,\nu}^{T\bar{T}}$
and v$_{\mu,\nu}^{TT}$ are respectively the elements of the overlap matrix,
kinetic operator and the external potential $v_{ext}({\bf r})$ due
to the nuclei.
The matrix elements $J_{\mu,\nu}^{TT}$ and $K_{\mu,\nu}^{TT}$
are associated with the Coulomb operator, $ v_{H}^{L}[\rho({\bf r})]$
and  the exchange-correlation potential, $v_{xc}^{L}[\rho({\bf r})]$, respectively.
The matrix $H_{DKS}$ depends, through $\rho$ in
$v_{xc}^{L}[\rho({\bf r})]$ and $v_{H}^{L}[\rho({\bf r})]$,
on the canonical spinor-orbitals produced by its diagonalization, so that the solution (${\bf c^{T}}$) can not be obtained in a single step, but in a SCF procedure.

The most expensive  steps for a 4c-DKS calculation are the evaluation of the Coulomb and exchange correlation matrix elements (Eq.\ref{jelement} and Eq.\ref{kelement}). We have shown that these computational tasks can be  reduced  by using the density fitting approach\cite{BelpB16_20, belpassi11_12368,belpassi2020bertha}.
Substantially, from a $O(N^2)$ scaling for the the total charge density, written in the terms of overlap spinor density (Eq.\ref{totchargerho}), 
one can reduce the scaling factor to linear using auxiliary basis functions ($f_t(\textbf{r})$):
\begin{equation}
    \tilde\rho(\textbf{r}) = \sum_t d_t f_t(\textbf{r}),
\end{equation}

The density fitting scheme relies on selecting an auxiliary basis set, \( \{ f_t \} \), and determining the coefficients ($d_t$) that best approximate the total electron density. This approach not only simplifies the construction of the required matrices but also avoids the evaluation of four-center in favour of three-center two electron integrals with a minimal loss of accuracy.
The coefficients $d_t$ are chosen to minimize the error between the fitted density $\tilde\rho(\mathbf{r})$ and the true density $\rho(\mathbf{r})$ using the Coulomb metric ($g(r-r')=\frac{1}{|r-r'|}$):
\begin{equation}\label{error}
\Delta =\frac{1}{2}\int d\mathbf{r}\int
d\mathbf{r'}{[\rho(\mathbf{r})-\tilde{\rho}(\mathbf{r})]g(r-r')[\rho(\mathbf{r'})-\tilde{\rho}(\mathbf{r'})]}{}.
\end{equation}
This approach is widely used in non-relativistic cases and has been extensively studied\cite{Dunlap:1990}. The variational Coulomb fitting method, often simply referred  to  Coulomb fitting, ensures a positive-definite metric, leading to a minimization procedure of $\Delta$ which is bound from below
and gives a linear system for the fitting coefficients ${\bf d}$:
\begin{equation}
{\bf A\,d}={\bf v},
\label{eq:Adv}
\end{equation}
where $A_{st}$ represents the Coulomb matrix of auxiliary functions:
\begin{equation}
A_{st}=\left <f_s||f_t\right>\equiv
\int f_s({\bf r})\frac{1}{|{\bf r}-{\bf r}'|}f_t({\bf r}')\,d{\bf r}
d{\bf r}'.
\end{equation}
The vector $\bf v$ projects the electrostatic Coulomb potential onto the fitting functions:
\begin{equation}
v_s  = \left< f_s||\rho \right> =
\sum_{\mu\nu}
\left( I^{LL}_{s,\mu\nu}D^{LL}_{\mu\nu}+
I^{SS}_{s,\mu\nu}D^{SS}_{\mu\nu}  \right),
\end{equation}
which, as we can see, can be expressed in terms of the density matrix elements $D^{LL}_{\mu\nu}$ and of the 3-center two-electron repulsion integrals $I_{s,\mu \nu}^{TT}$ 
\begin{equation}
I_{s,\mu \nu}^{TT} =  \left < f_s|| \rho^{TT}_{\mu \nu} \right>.
\end{equation}
involving both the fitting functions $f_s$ and the charge overlap terms $ \rho^{TT}_{\mu \nu}$.

The Coulomb matrix, therefore, can be expressed as:
\begin{equation}\label{Jfit}
\tilde J_{\mu\nu}^{TT} =
\left < \tilde \rho|| \rho^{TT}_{\mu \nu} \right > =
\sum_{t=1}^{N_{\mbox{\scriptsize aux}}} I^{TT}_{t,\mu\nu}d_t.
\end{equation}

Once the density fitting approach is adopted to compute
the Coulomb contribution,  the
computation bottleneck moves on to the evaluation of the exchange-correlation matrix.
The original idea of directly using the density expansion obtained from the 
Coulomb fitting into the exchange–correlation functional is
relatively old~\cite{Dunlap:1990,Laikov:1997}  and a number of modern effective implementations
have been proposed.  We follow the same scheme proposed by K\"{o}ster et al.~\cite{Koster:2004}.

Therefore, after some derivations (see Refs.\cite{Koster:2004,belpassi11_12368} for details), we can
obtain the  exchange-correlation matrix contribution, similar to Eq.\ref{eq:Adv}, solving a linear system for a new set of coefficients $\{z\}$
\begin{equation}\label{eq:Azw}
{\bf A\,z}={\bf w},
\end{equation}
where 
\begin{equation}\label{eq:w}
w_s  = \left< v_{xc}[\tilde \rho]|f_s \right>.
\end{equation}
The above integral is evaluated numerically.

The total ($J_{\mu \nu}^{TT}+\tilde K_{\mu \nu}^{TT}$) matrix contribution can be computed in a single step:
\begin{equation}\label{singlestepC-XC}
\tilde J_{\mu \nu}^{TT}+\tilde K_{\mu \nu}^{TT}=  
\sum_{t=1}^{N_s} I^{TT}_{t,\mu\nu}(d_t+z_t).
\end{equation}
in terms of three-index two electron integrals ($I^{TT}_{t,\mu\nu}$).
The application of such method achieves high-accuracy numerical integration and scales  as $N_s \cdot N_{grid}$ instead of $N^2 \cdot N_{grid}$.
The effectiveness of this approach
is further enhanced by using, as fitting functions, primitive
HGTF that are grouped together in sets sharing the same
exponents.
The sets are formed so that to an auxiliary function
of given angular momentum are associated all the functions
of smaller angular momentum. For instance, a $d$ auxiliary
function set contains ten primitive Hermite Gaussians, one $s$,
three $p$, and six $d$ functions all with the same exponent. This
scheme allows us to use efficient  recurrence relations of
Hermite polynomials in the computation of two-electron
integrals. Analogous schemes have been adopted in Ref.s\citenum{Koster:2004,Koster:2025}.

\section{Automatic generation of ABS}\label{DFBERTHA}
The fundamental component of the density fitting scheme is the possibility of having an accurate auxiliary basis set.
As already mentioned, there are two main approaches to obtain an ABS: 1) through an optimization procedure based on atomic or molecular calculations; or 2) through an automatic generation procedure starting from a principal basis set.
The algorithm described here automatically generates an ABS for a given principal G-spinor basis set.
The algorithm presented below does not pretend to be optimal, but as will be shown in the next section, it allows us to generate ABSs that are surprisingly accurate for the entire periodic table.
The algorithm is specific for the ABS structure we use in BERTHA, which, as already stated, is formed of primitive  HGTFs grouped in sets sharing the same exponents.
 Based on this definition,  only two parameters are required to define each group of HGTFs: the exponent ($\alpha$) and the associated angular momentum value ($l_{fitt}$). The group is formed in such a way that all functions with an angular momentum smaller than $l_{fitt}$ (in Eq.\ref{HGTFgeneraleq} the indexes $i$, $j$, $k$ assume all values with the condition that $i+j+k\leq l_{fitt}$) belongs to the auxiliary function with the same exponent,$\alpha$. The total number of HGTFs for a given exponent, $\alpha$, are given by
\begin{equation}
    \text{N}_{\alpha} = \frac{(l_{fitt}+3)(l_{fitt}+2)(l_{fitt}+1)}{6}
\end{equation}
For example, for a value of $l_{fitt}=3$ we have a total of twenty 
primitive Hermite-Gaussians, one $s$,
three $p$, six $d$ and ten $f$ functions, all of which have the same exponent ($\alpha$).
The assignment of this angular momentum label is a crucial step in the construction of the basis set. It assumes particular importance for heavy elements which contain electrons with high angular momentum values. An effective algorithm should select value of $l_{fitt}$ to achieve an optimal trade-off between computational cost and accuracy.

To generate the ABS exponents, we used an algorithm developed several years ago Ref.~\cite{Koster2018}. It was originally used to generate the ABS for elements up to the $3d$ elements (from H to Kr) in non-relativistic DFT calculations using LDA or GGA exchange-correlation functionals. In this context, the ABSs generated with this algorithm are referred to as GEN-A$n$ (see Ref. \cite{Calamicini:2007} for details). 
This algorithm is based on an even-tempered method and uses as input the primitive
Gaussian exponents of the principal basis set (the smallest $\beta_{min}$ and the largest, $\beta_{max}$).
The latter values, together with the integer parameter $n$, determine the total number of exponents ($N_{exp}$) that must be generated, according to the following formula:
\begin{equation}\label{NauxDEMON2k}
    \mathrm{N_{exp}} = \mathrm{Int} \left( \frac{\ln \left( \beta_{max} / \beta_{min} \right) } {\ln \left( 6-n \right) } + 0.5 \right)
\end{equation}
Here $n$ can be assigned an integer value of 2, 3 or 4, which increases the total number of exponents to be generated accordingly.

The tightest seed for the geometric progression ($\alpha_{0}$) is defined as follows: 
\begin{equation}
    \alpha_{0} = 2 \beta_{\text{min}}(6-n)^{(N-1)}
\end{equation}
Then, the first exponent of the fitting set is given by
\begin{equation}
   \alpha_{1} = \left( 1+\frac{n}{12-n} \right)\alpha_{0}
\end{equation}
and the second one reads as
\begin{equation}
    \alpha_{2} = \left( \frac{\alpha_{0}}{6-n} \right).
\end{equation}
Finally, we just need to apply the following geometric progression until all the required $N_{exp}$ exponents have been generated
\begin{equation}
\alpha_{i+1} = \left( \frac{\alpha_{i}}{6-n} \right).
\end{equation}
Once we have the set of exponents (corresponding to a certain value of $n$),
we need to assign the angular part. This means that we need to set
 a value of $l_{fitt}$ to each exponent. This is an important step in our algorithm as it greatly affects both the accuracy and the computational cost. We use a very simple method here.
As already mentioned, G-spinors can be constructed from a conventional, non-relativistic definition of the basis functions, which are given as terms of Gaussian exponents ($\beta$) with respect to the orbital angular momenta $l$.
Based on this  criterion, we double the values of the exponents for each value of angular momentum and define blocks of exponents in the range from $2\beta_{min,l}$ to $2\beta_{max,l}$.
This gives us a simple procedure for assigning an angular momentum value to our group of suitable exponents.
In particular, if an exponent $\alpha_i$ of our ABS is $2\beta_{min,l}\leq \alpha_i \leq 2\beta_{max,l}$, we temporarily assign an angular value ($l_{temp}$) corresponding to that of the principal basis set ($l$).
In cases where an exponent falls into multiple blocks with different $l$, we assign the value with the largest $l$ to the ABS exponent. 
This strategy is useful because it extracts the angular momentum information from the principal basis set, but  using these values of $l_{temp}$ directly in the ABS definition,  gives inaccurate results. This is not surprising given the low angular flexibility.
Indeed, the exact expansion of two spherical harmonics with angular quantum number $l_1$ and $l_2$ involve a linear combination 
of spherical harmonics of order $L$ weighted by Clebsch-Gordan coefficients
\begin{equation}
    Y^{m_1}_{l_1}Y^{m_2}_{l_2} = \sum^{l_1+l_2}_{L = |l_1-l_2|} C(l_1,l_2,L,m_1,m_2)Y^{m_1+m_2}_L
    \label{eq:expansion}
\end{equation}
with $L$-values in the range from $|l_1-l_2|$ to $l_1+l_2$.
This means that in order to fully capture the overlap expansion, combinations of angular values up to $l_1+l_2$ would have to be considered. Since the small component of the G spinors is generated via the kinetic equilibrium prescription (which performs a spatial derivative), the expansion would even lead to a higher order spherical harmonic (up to $L=|l_1+l_2+2|$), see Ref.\cite{grant07}. This simple analysis shows that if the generated exponent falls into the $f$-shell (l = 3), we would have to assign $l_{fitt} = l_1 + l_2 +2 = l_1*2 +2 = 8$.
It is easy to see that if large polarised basis sets were used for heavy elements, the number of matching functions would quickly explode.
Fortunately, in the Dirac-Kohn-Sham method, which uses local (LDA) or semi-local (GGA) exchange correlation functions, the G-spinor overlap density is always contracted with the one-particle density matrix (see Eq. \ref{totchargerho}). This greatly reduces the angular dependence of the quantities involved  and it is actually not necessary to include such a strong angular dependence in ABS as equation \ref{eq:expansion} would suggest.
The situation is different if we use density fitting in calculations with exchange-correlation functions that contain a certain amount of exact Hartree-Fock 
exchange (see Ref.s\citenum{Koster:2025,bagel:2013}).
We have pragmatically found that, starting from the $l_{temp}$ values defined above, it is sufficient to increase them by one or two units to obtain reasonably accurate results.
This prompted us to define two different cases
\begin{eqnarray}
 l_{fitt}=l_{temp}+1 && l_{fitt}=l_{temp}+2
\end{eqnarray}
For each of these cases, we have generated two different ABS by setting $n = 2$ and $n = 3$, resulting in four different ABS, namely GEN-n2-v1 ($n=2$, $l_{fitt}=l_{temp}+1$), GEN-n2-v2 ($n=2$, $l_{fitt}=l_{temp}+2$),
GEN-n3-v1 ($n=3$, $l_{fitt}=l_{temp}+1$) and GEN-n3-v2 ($n=3$, $l_{fitt}=l_{temp}+2$).

To summarise, to fully generate the ABS we need to retrieve the following information from the main base set:
\begin{itemize}
 \item the highest ($\beta_{max}$) and lowest ($\beta_{min}$) exponents
 \item the highest and lowest exponents for each atomic shell present in the principal basis (necessary for the assignment of the $l_{temp}$ and then the $l_{fitt}$ value)
\end{itemize}

The procedure was implemented in Python and is freely available\cite{fitgen}.

\section{Computational details}

In the present section we will give some details about the computational procedure we have adopted to efficiently automatize the entire workflow which allowed us to generate different ABSs and efficiently test their accuracy on a large set of molecules. 

\begin{figure}[H]
    \centering
    \captionsetup{justification=centering}
    \includegraphics[width=1\textwidth]{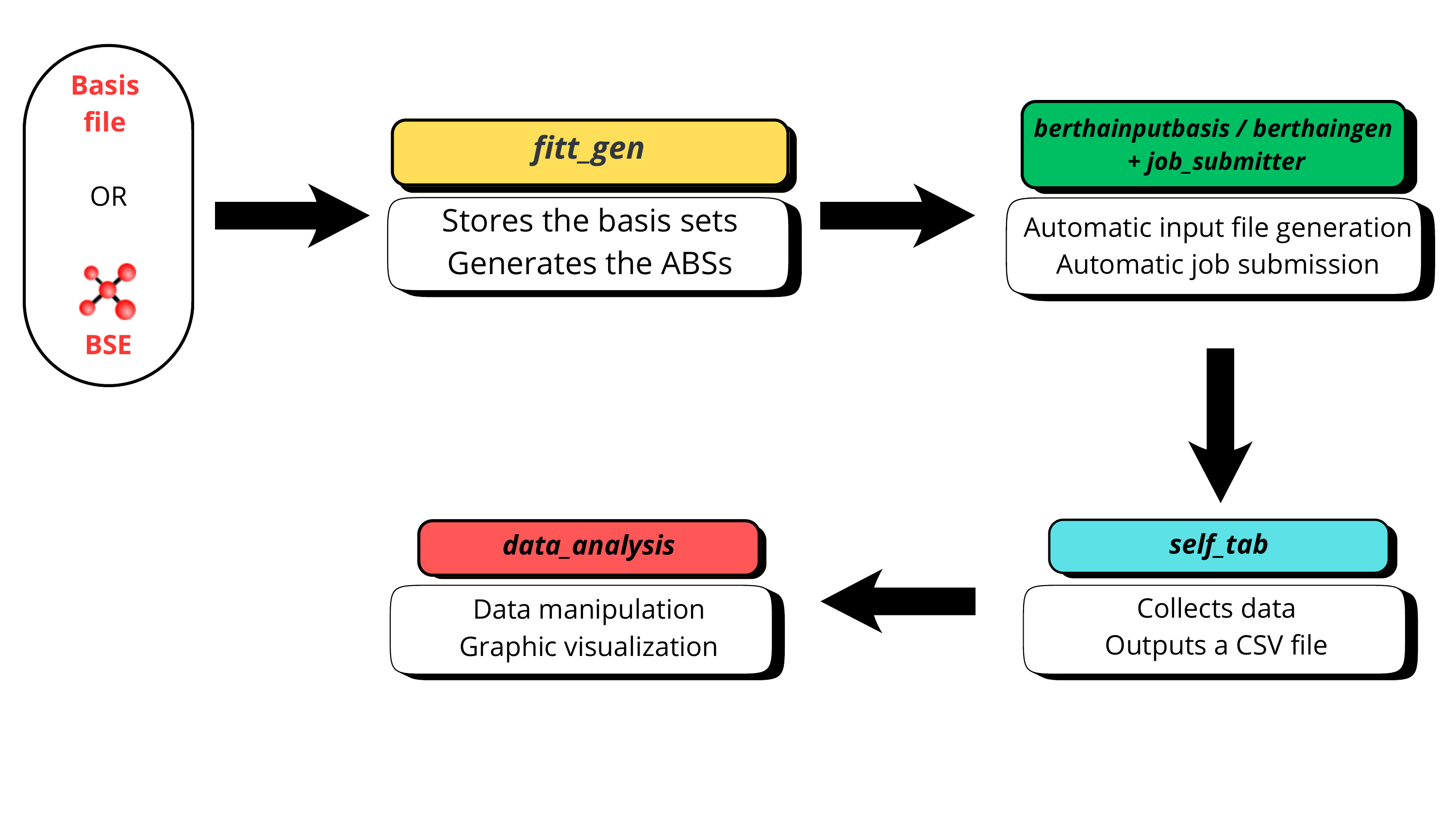}
    \caption{Workflow diagram of the implemented procedure to i)  generate ABS, ii) construct BERTHA inputs and run them, iii) collect data in CVS file and iv) data analysis.}
    \label{workflow}
\end{figure}

The workflow is depicted in Figure~\ref{workflow}. It begins 
by selecting a principal G-spinor basis set, which is used to generate the auxiliary basis set (performed by the Python script \textit{fitt\_gen})\cite{fitgen}.
 The cited Python script can also be used to directly download any desired basis set from the Basis Set Exchange (BSE) \cite{pritchard2019new} website.
The algorithm described in the previous section will then automatically generate
 the auxiliary basis sets for each atom. The generated basis and fitting sets are stored in a JSON file via the \texttt{berthainputbasis} 
 Python modules \cite{berthainputbasis}.  \texttt{berthainputbasis} is a Python module that has been developed to collect and organize all G-spinor basis sets and auxiliary basis sets in a single JSON file. The JSON file,
 together with the molecular geometry files (in XYZ format), is then used by the \texttt{ berthaingen} \cite{berthaingen} module to construct the input file to perform the DKS calculation in BERTHA. While the \texttt{berthaingen} can be used as a stand-alone Python script, it can also be imported as a proper Python module. 
Subsequently, the \textit{job\_submitter}  \cite{fitgen} Bash script, that automates processing of input generation and calculations submission, has been developed. 
The \textit{self\_tab} \cite{fitgen} program, which serves as a CSV database generator, collects all the the data and parses them into a single CSV file. 
Finally, the \textit{data\_analysis} Jupyter notebook streamlines statistical analysis 
by performing data manipulation and visualization.
This workflow enhances automation, simplifies the data analysis and facilitates auxiliary 
function development across a large set of molecular structures.  

\subsection{Molecular dataset}\label{dataset}
Here, we briefly describe the molecular dataset used to verify the accuracy of our  automatically generated ABS. As mentioned earlier, due to the all-electron nature of our DKS calculations, our goal was to develop a method accurate enough for the entire periodic table, including the heavy and superheavy elements.
This requires the testing of our automatic scheme on a large molecular data set representative of all groups and periods of the periodic table.
Pollak and Weigend \cite{PolWeigSET} published a large molecular data set that included systems containing almost all elements of the periodic table except for the 7th period elements. We used this molecular data set as a starting point, 
removing those molecules with a multi configurational character which fail to converge in our DKS calculations.
All the removed molecules contain $f$ elements.
For this reason, the molecules with $f$ elements in the data set are limited to a few closed-shell molecules, namely:
$CeH_2$, $DyF_2$, $La_2O_3$, $LaF_3$, $LaI_3$ and $UF_6$.
The molecular data set has been further expanded to include elements from the 7th period and noble gases that were both missing in the original dataset.
The molecules that we choose for representing the 7th period are those of the 6th period 
where the heavy element is replaced by the corresponding element of the 7th period. The geometries for these systems were optimized using the ADF software \cite{ADF2001} at ZORA level including the spin-orbit coupling in combination with the BLYP exchange correlation functional and TZ2P basis set.
In order to include noble gases, we use the complexes $OBeX$ ($X = He, Ne, Ar, Kr, Xe, Rn, Og$). In these systems the $OBeX$ molecule, with its very large dipole moment\cite{jpcl_belpassi:2017}, allows to strongly polarize the noble gases and thus is a perfect test for the ABS in these systems.   
The final dataset included 286 molecules and covered the entire periodic table, with the exception of the $f$ block, which, as already mentioned, are only sparsely represented due to their multireference character. The final version of the molecular dataset 
includes all molecular geometries in XYZ format and  is freely available at the public repository \cite{berthatestbench}.

\section{Results and discussion}

The density-fitting scheme implemented in the DKS module of BERTHA ensures that the Coulomb energy in Eq.\ref{error} is approximated from above. We have already shown many years ago that for molecules containing heavy elements, very accurate results can be obtained with
 error in the Coulomb energy below $10^{-4}-10^{-5}$ hartree (of the order of $\mu$hartree {\it per electron}). This error is of the same order of magnitude as in the non-relativistic Kohn-Sham implementations\cite{EICHKORN1995283,manby:2001,Lehtola1,Lehtola2} that use density fitting for molecules with light elements or using pseudopotentials.
 As already mentioned, in our case the fitted density is also used for the evaluation of the exchange correlation term. This leads to errors in the total energy that are typically one or two orders of magnitude larger, but we have clearly shown \cite{PhysRevB.77.233403} (see also
K\"{o}ster et al. in the non-relativistic context\cite{Koster:2004}) that all chemically relevant quantities (binding energies, harmonic frequencies, geometries and properties) can be obtained in excellent agreement with exact calculations (without density fitting).
Very simple schemes can be used (see Ref.\cite{PhysRevB.77.233403}) to increase the accuracy of the total energies. In the Supporting information we report an explicit example for the Au$_2$ molecule.
For our purposes here, we evaluate the accuracy of our ABS using the error in the Coulomb energy (Eq.\ref{error}).

\begin{figure}[h]
    \centering
    \begin{subfigure}{0.90\textwidth}
    \captionsetup{justification=centering}
    \includegraphics[width=\linewidth]{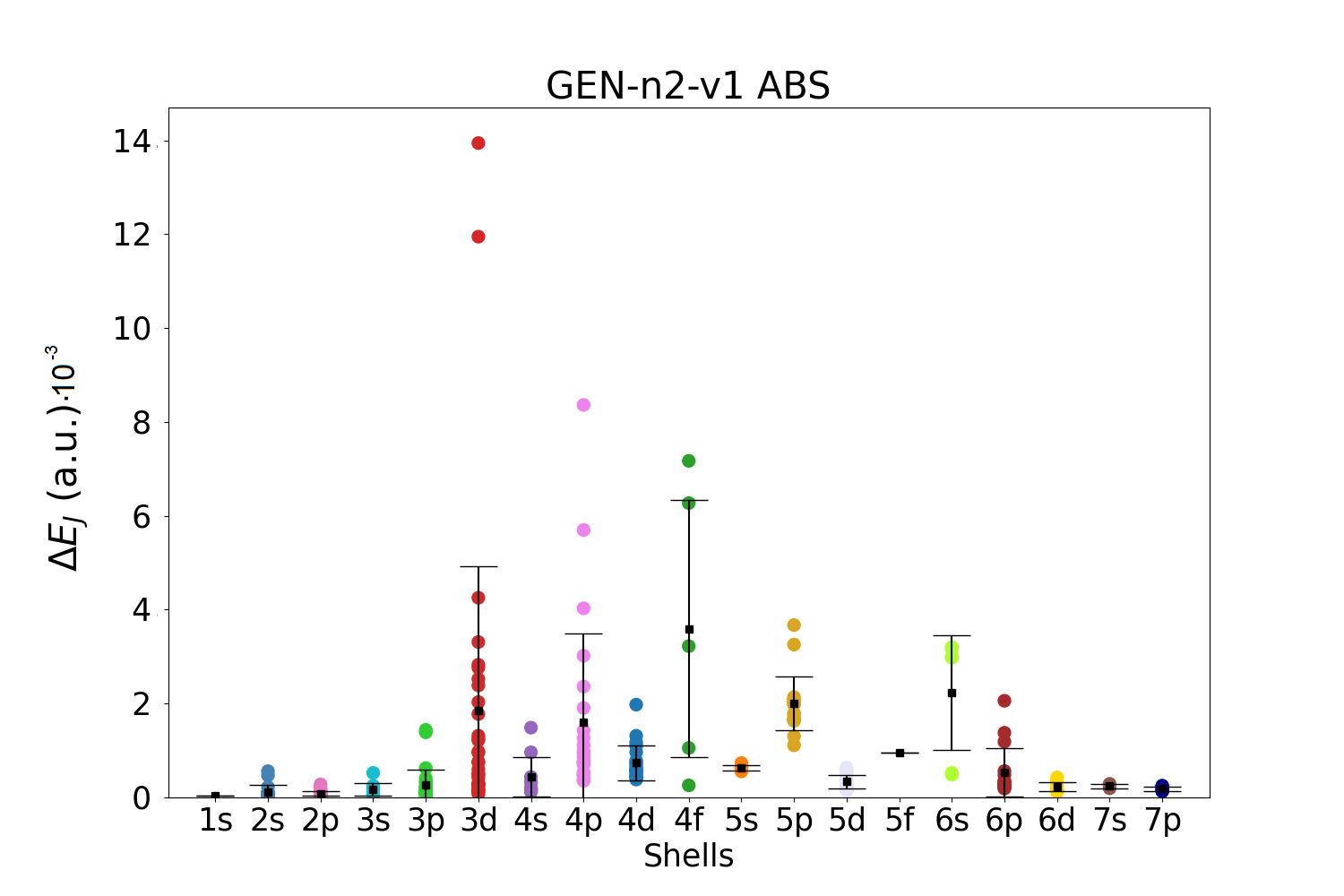}
    \caption{ $\Delta E_J$ (a.u.)}
    \end{subfigure}
    \hfill
    \begin{subfigure}{0.90\textwidth}
    \includegraphics[width=\linewidth]{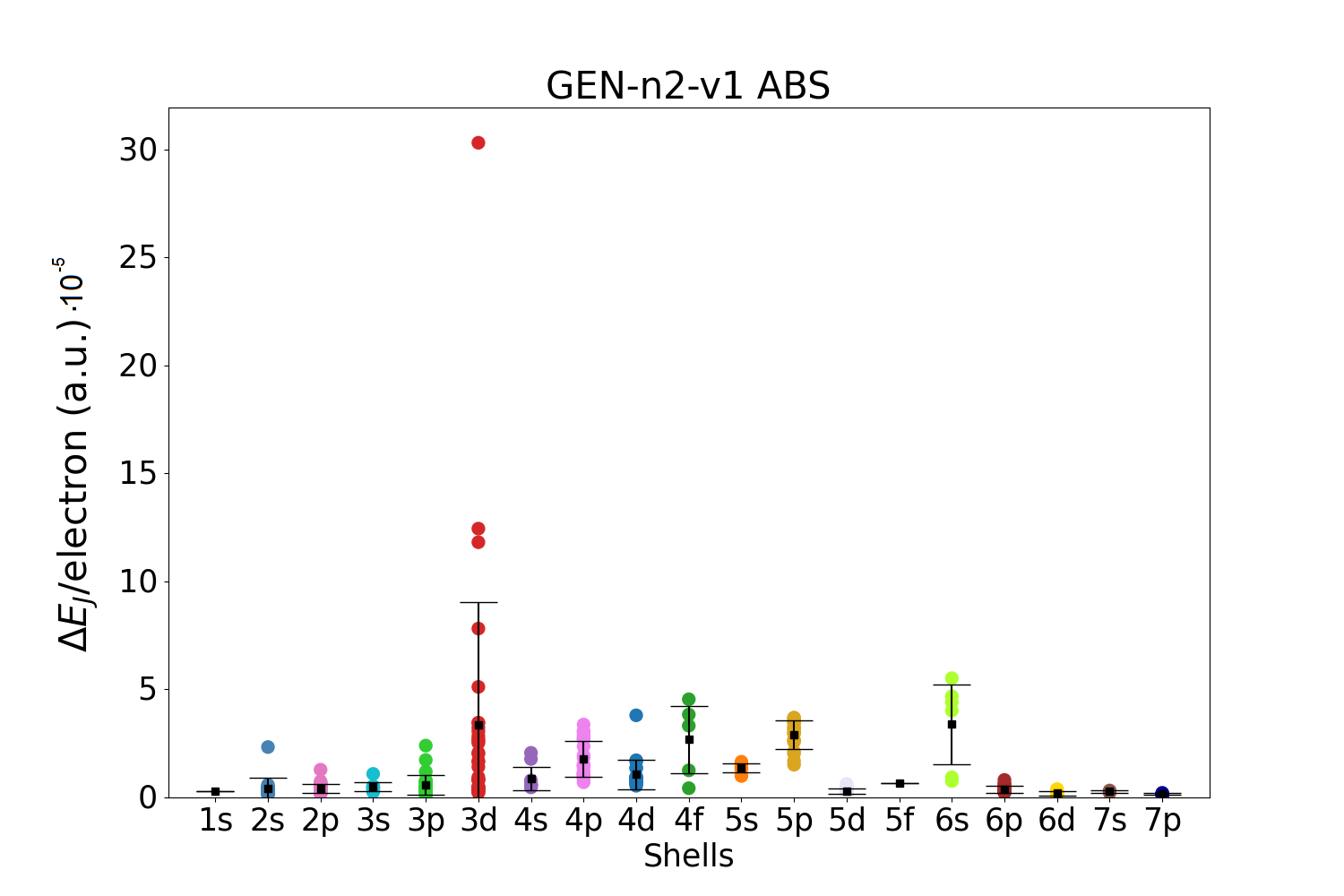}
    \caption{$\Delta E_J$ {\it per electron}  (a.u.)}
    \end{subfigure}
    \caption{Coulomb energy error ($\Delta E_J$, panel a) and Coulomb energy error per electron ($\Delta E_J/per\ electron$, panel b) for all molecules in the data set. Mean error (black squares) and standard deviation have been also reported. Data obtained using Dyall.v2z basis set in combination with the automated generated \texttt{GEN-n2-v1} ABS.}
    \label{n2v1} 
\end{figure}

\begin{figure}[h]
    \centering
    \begin{subfigure}{0.90\textwidth}
    \centering
    \captionsetup{justification=centering}
    \includegraphics[width=\linewidth]{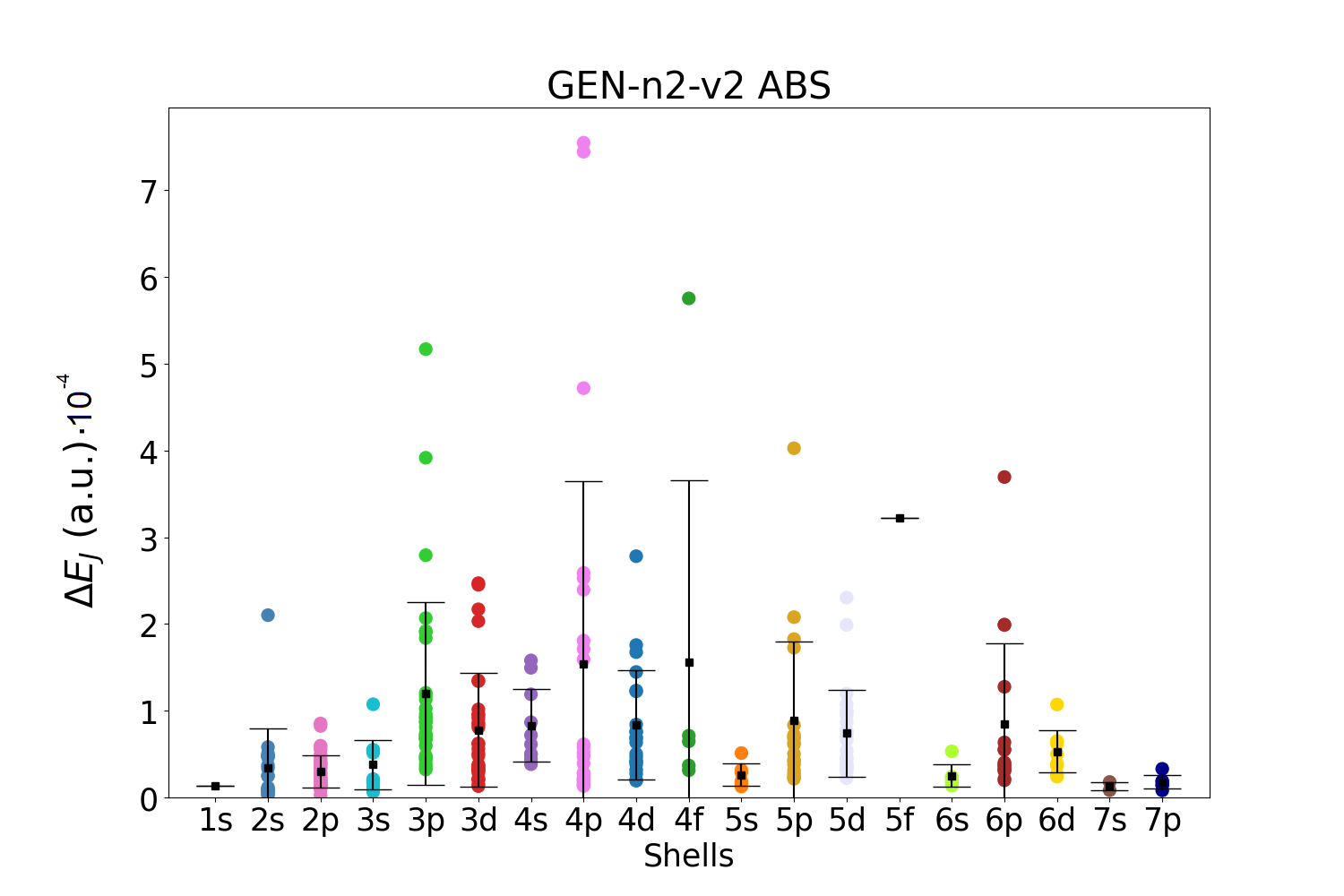}
    \caption{ $\Delta E_J$ (a.u.)}
    
    \end{subfigure}
    \hfill
    \begin{subfigure}{0.90\textwidth}
    \centering
    \includegraphics[width=\linewidth]{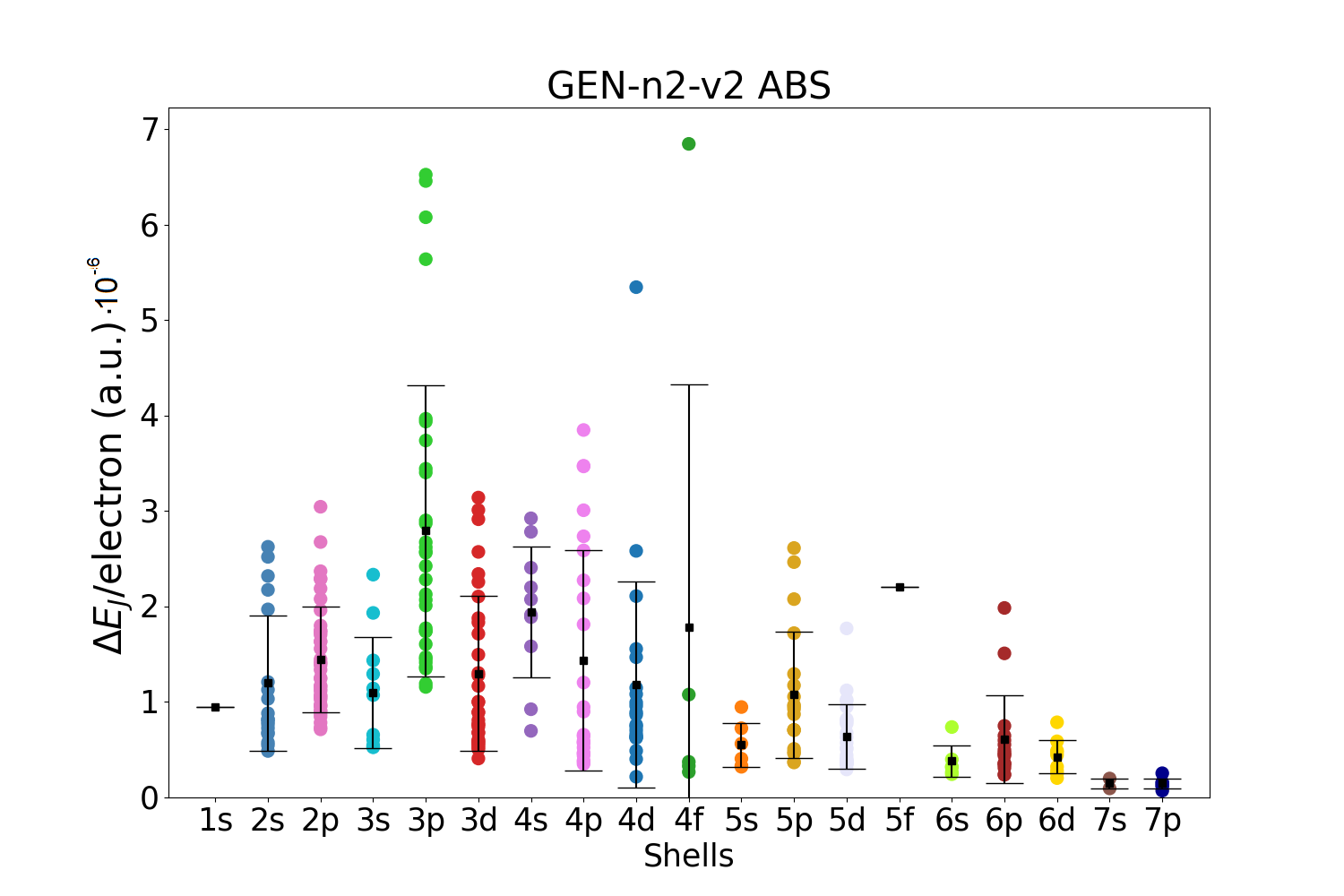}
    \caption{$\Delta E_J$ {\it per electron}  (a.u.)}
    \end{subfigure}
    \caption{Coulomb energy error ($\Delta E_J$, panel a) and Coulomb energy error per electron ($\Delta E_J/per\ electron$, panel b) for all molecules in the data set. Mean error (black squares) and standard deviation have been also reported. Data obtained using Dyall.v2z basis set in combination with the automated generated \texttt{GEN-n2-v2} ABS.}\label{n2v2}
\end{figure}

\begin{figure}[h]
\centering
    \begin{subfigure}{0.90\textwidth}
    \captionsetup{justification=centering}
    \includegraphics[width=\linewidth]{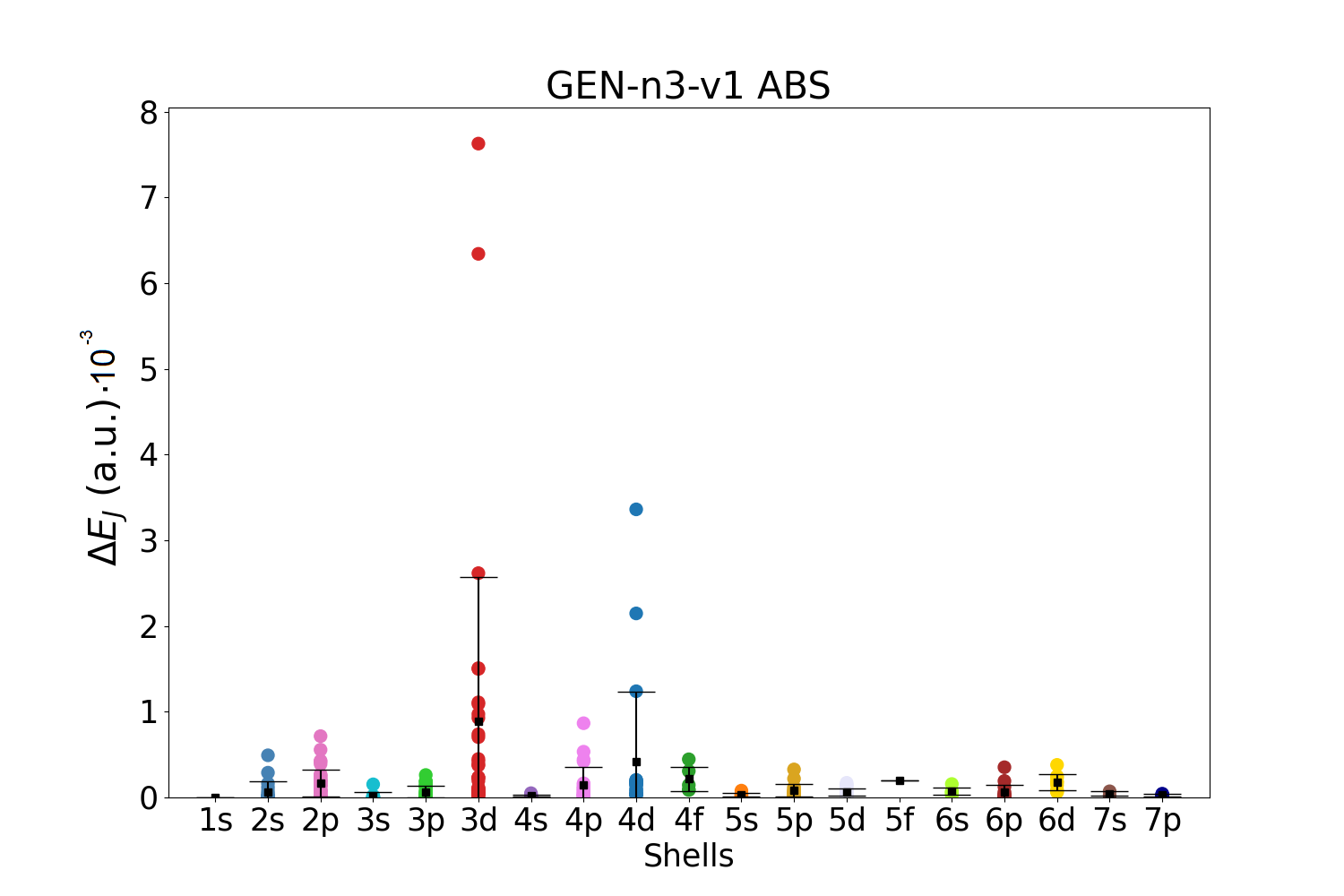}
    \caption{ $\Delta E_J$ (a.u.)}
 
    \end{subfigure}
    \hfill
    \begin{subfigure}{0.90\textwidth}
    \includegraphics[width=\linewidth]{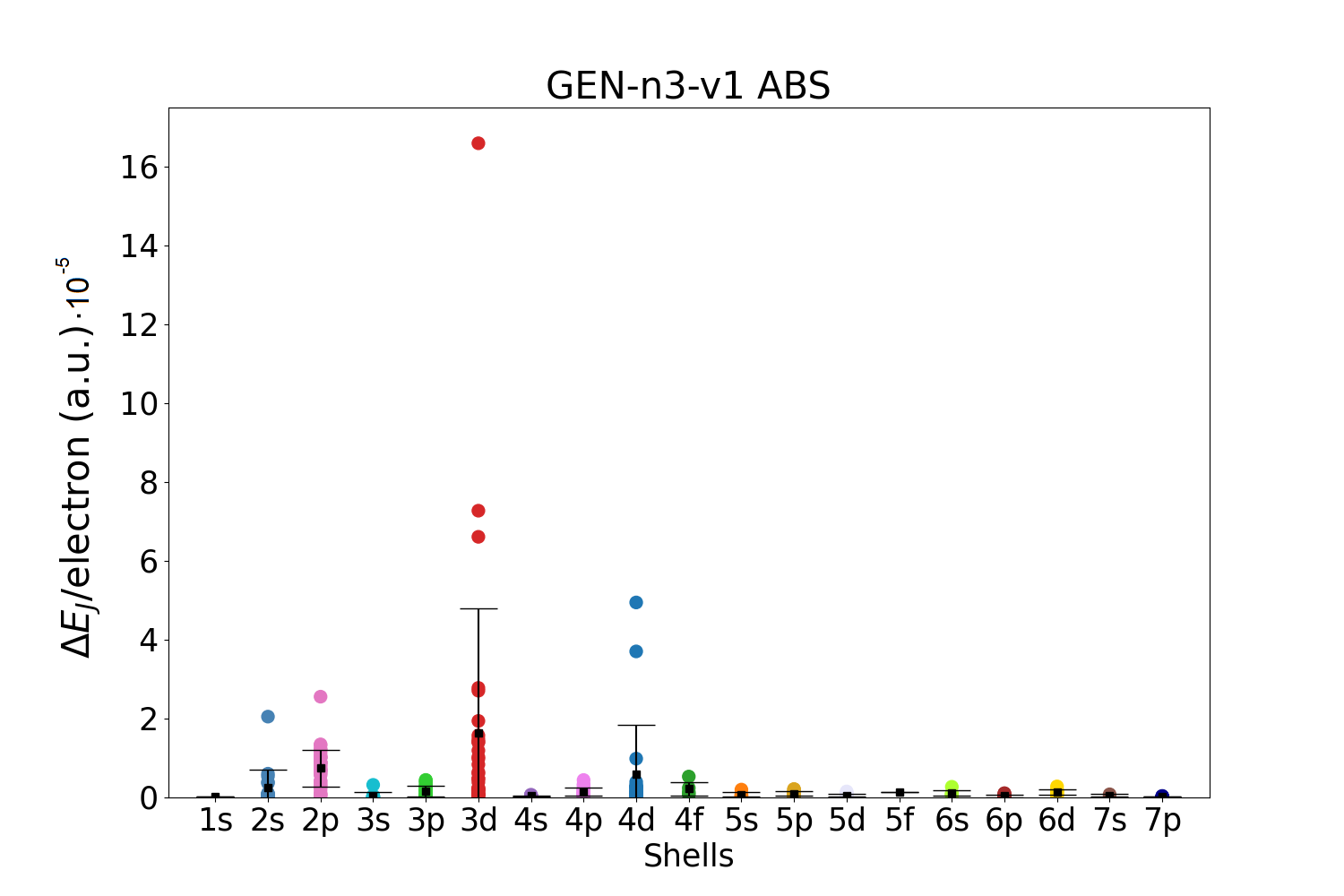}
    \caption{$\Delta E_J$ {\it per electron}  (a.u.)}
    \end{subfigure}
    \caption{ Coulomb energy error ($\Delta E_J$, panel a) and Coulomb energy error per electron ($\Delta E_J/per\ electron$, panel b) for all molecules in the data set. Mean error (black squares) and standard deviation have been also reported. Data obtained using Dyall.v2z basis set in combination with the automated generated \texttt{GEN-n3-v1} ABS.}   
    \label{n3v1}
\end{figure}

\begin{figure}[h]
    \centering
    \begin{subfigure}{0.90\textwidth}
    \captionsetup{justification=centering}
    \includegraphics[width=\linewidth]{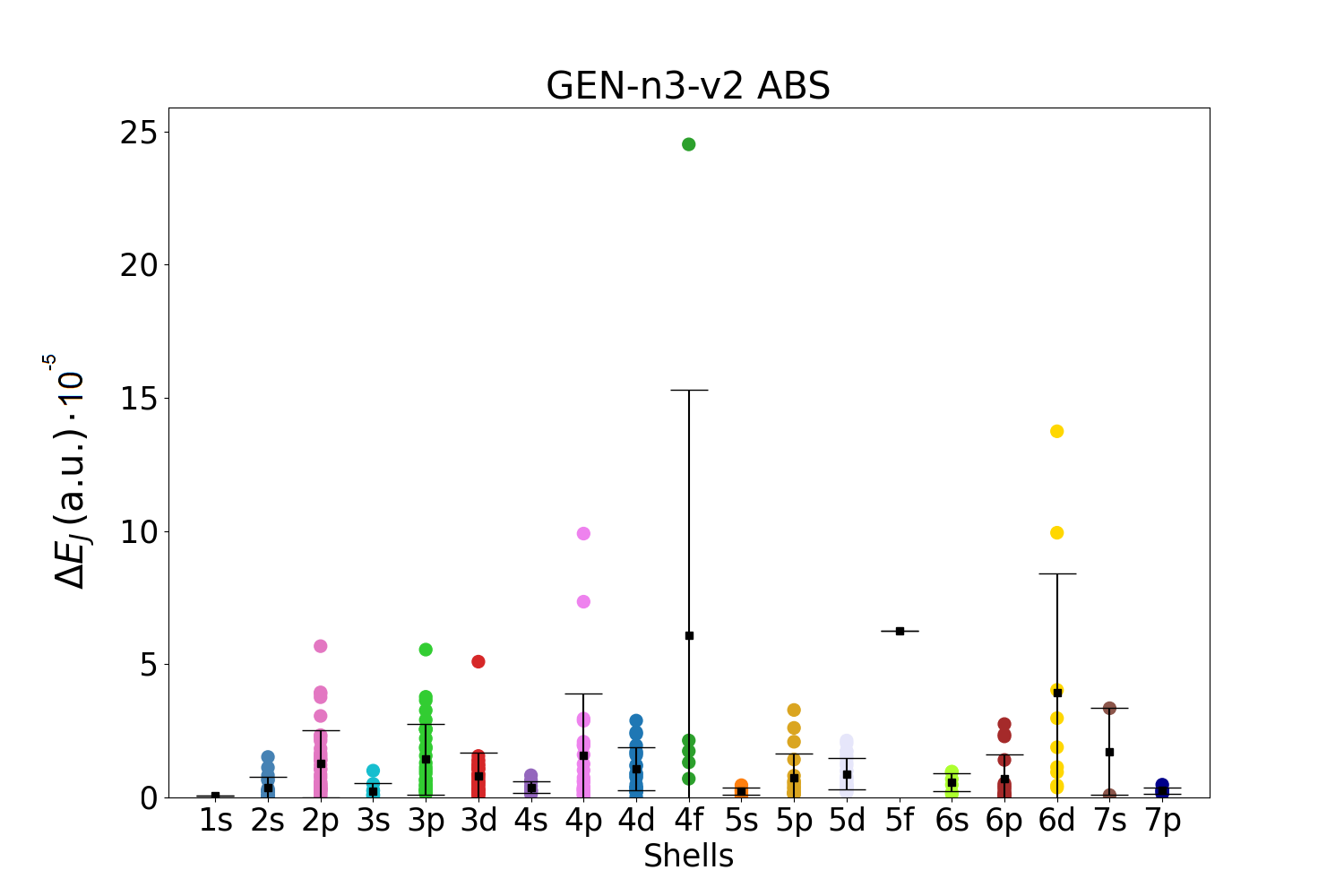}
    \caption{ $\Delta E_J$ (a.u.)}.

    \end{subfigure}
    \hfill
    \begin{subfigure}{0.90\textwidth}
    \includegraphics[width=\linewidth]{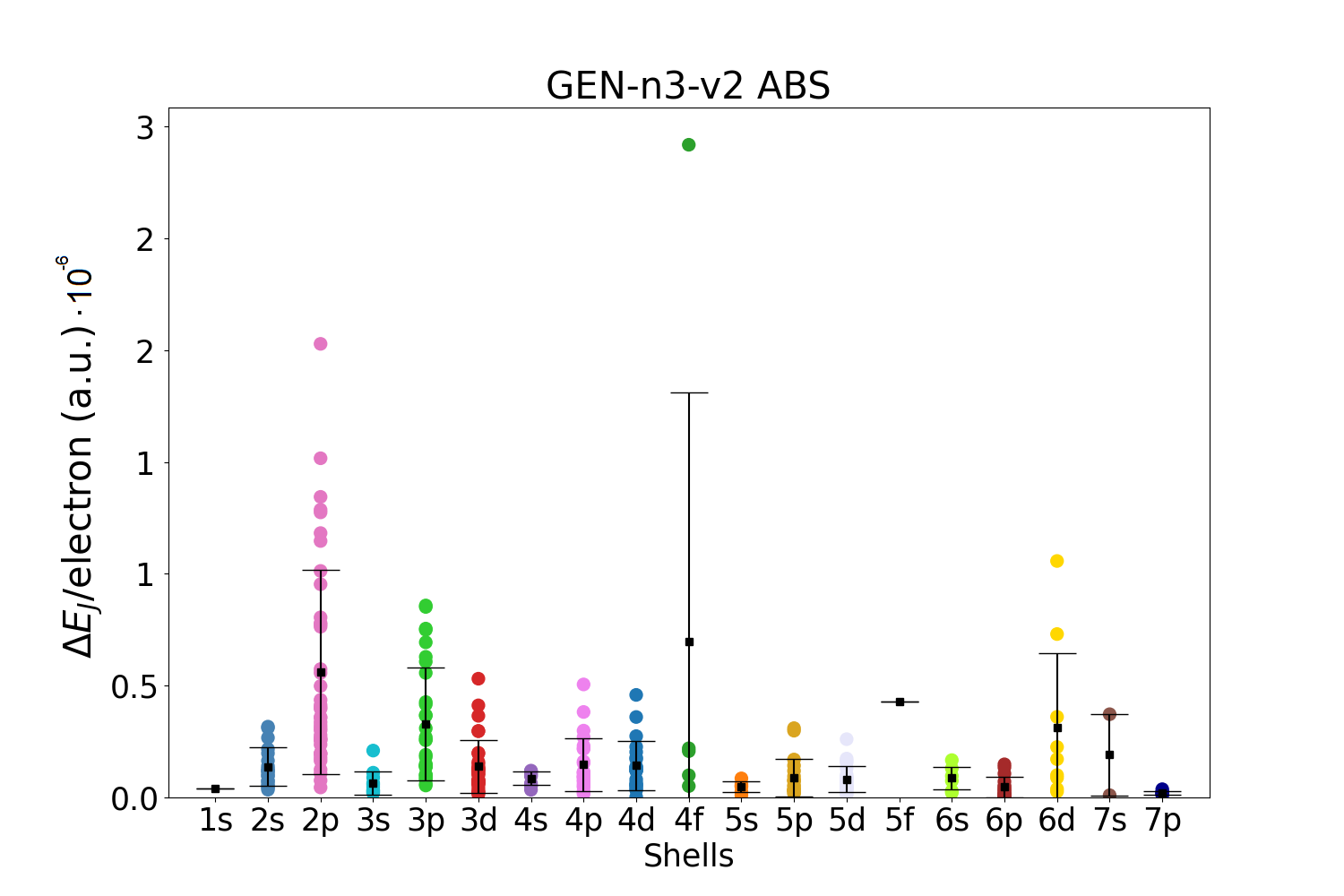}
    \caption{$\Delta E_J$ {\it per electron}  (a.u.)}
    \end{subfigure}
    \caption{Coulomb energy error ($\Delta E_J$, panel a) and Coulomb energy error per electron ($\Delta E_J/per\ electron$, panel b) for all molecules in the data set. Mean error (black squares) and standard deviation have been also reported. Data obtained using Dyall.v2z basis set in combination with the automated generated \texttt{GEN-n3-v2} ABS.}    \label{n3v2}
\end{figure}

We present the results obtained using Dyall's basis sets, namely \textit{dyall.v2z} and \textit{dyall.v3z} \cite{dyall_basis_sets}. These basis sets are explicitly designed for all electron-relativistic atomic and molecular structure calculations and are available for all elements in the periodic table, including superheavy atoms.
In conjunction with these principal basis sets, we tested the performance of four different auxiliary fitting basis sets.
The latter were automatically generated with two different values for $n$ and $l_{fitt}$, namely \texttt{GEN-n2-v1} ($n=2$, $l_{fitt}=l_{temp}+1$),
\texttt{GEN-n2-v2} ($n=2$, $l_{fitt}=l_{temp}+2$), \texttt{GEN-n3-v1} ($n=3$, $l_{fitt}=l_{temp}+1$) and \texttt{GEN-n3-v2} ($n=3$, $l_{fitt}=l_{temp}+2$).
Each auxiliary fitting basis set was tested with the molecular data set described in the previous section (286 molecules).
The results are shown in Figures~\ref{n2v1}, \ref{n2v2}, \ref{n3v1} and \ref{n3v2}. For each ABS, we give the absolute error in the total Coulomb energies ($\Delta E_J$, panel a) and the absolute error {\it per electron} ($\Delta E_J/electron$, panel b).
The mean value and the corresponding standard deviation are also shown in the Figures (for the numerical values, see Table S2 and S3 in SI).
\newpage

Our data show that the automatically generated ABS with the smaller value of $l_{fitt}$ (lower angular flexibility), namely \texttt{GEN-n2-v1} and \texttt{GEN-n3-v1}, are not entirely satisfactory. Although the mean value of $\Delta E_J$ is smaller than $3\cdot 10^{-3}$ and $9\cdot 10^{-4}$ hartree for all molecular groups (divided into shells), there are some molecular systems that exhibit errors up to $1.4\cdot 10^{-2}$ and $8\cdot 10^{-4}$ hartree, respectively, for \texttt{GEN-n2-v2} and \texttt{GEN-n3-v2}.
The accuracy increases significantly when ABS with higher angular flexibility are used (i.e. \texttt{GEN-n2-v2} and \texttt{GEN-n3-v2}).
In the case of \texttt{GEN-n2-v2} we obtained a mean value of $\Delta E_J$ and $\Delta E_J/electron$ always smaller than $8\cdot 10^{-4}$ hartree for all shells (in the worst case the error is of $1.6 \cdot 10^{-4}$ hartree for the $4f$ shell). Using the largest \texttt{GEN-n3-v2} basis set, we obtain an even higher accuracy, with an average error in Coulomb energy always below $6\cdot 10^{-5}$ (obtained for the 5f shell) and an absolute error below $2.5\cdot 10^{-4}$. In terms of error per electron, we can achieve an accuracy of $9\cdot 10^{-8}$ or $3\cdot 10^{-6}$ hartree. 

\begin{figure}[H]
    \centering
    \includegraphics[width=1\textwidth]{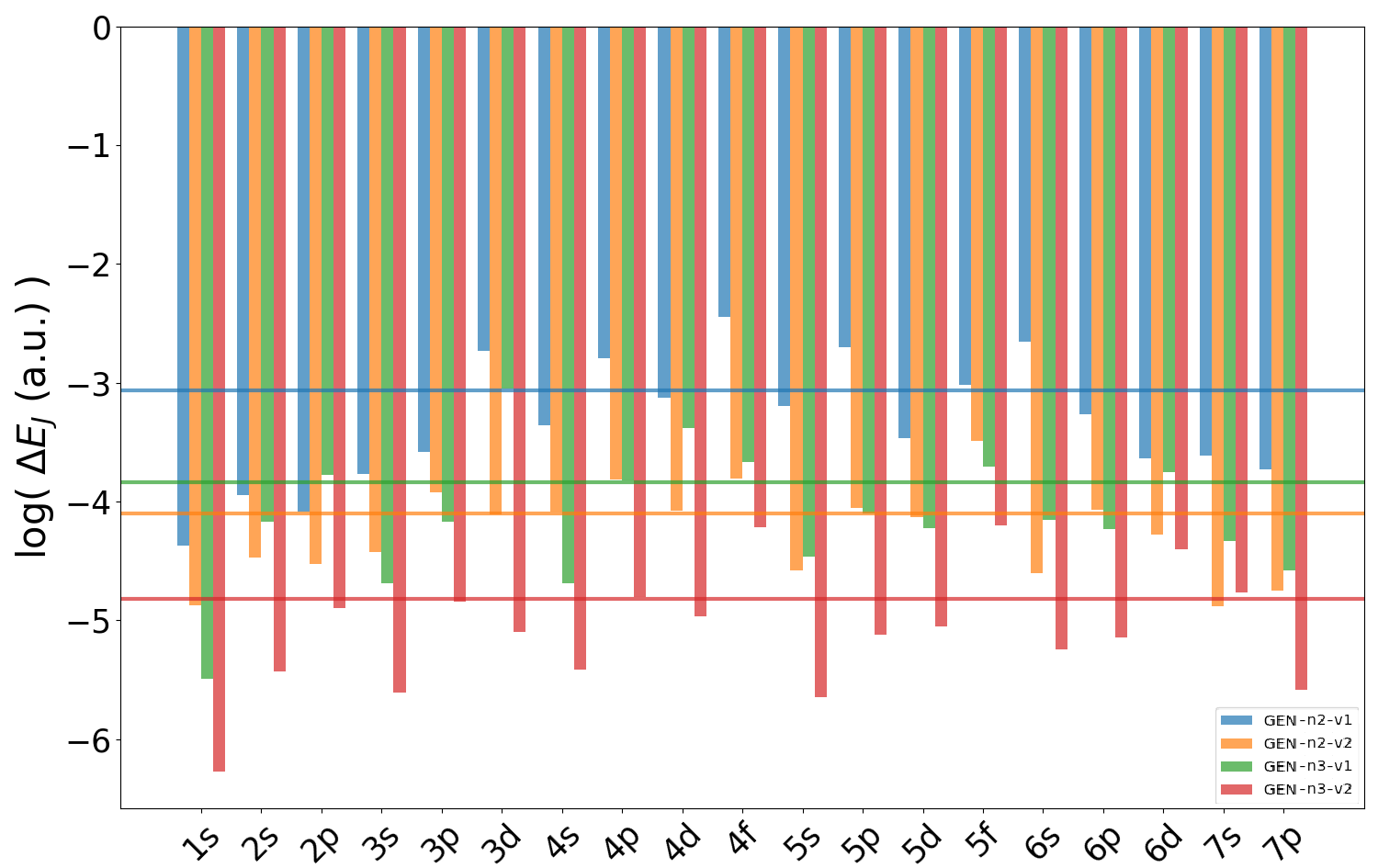}
    \caption{Mean value of $\Delta E_J$ in a.u. for each group of the molecular data base. The data are obtained using different ABSs, namely \texttt{GEN-n2-v1} (blue), \texttt{GEN-n2-v2} (yellow), \texttt{GEN-n3-v1} (green) and \texttt{GEN-n3-v2} (red). The horizontal lines with the same colors represent the mean value of $\Delta E_J$ for all molecular systems. The data are shown in logarithmic scale.}
    \label{histCoul}
\end{figure}

\begin{figure}[H]
    \centering
    \includegraphics[width=1\textwidth]{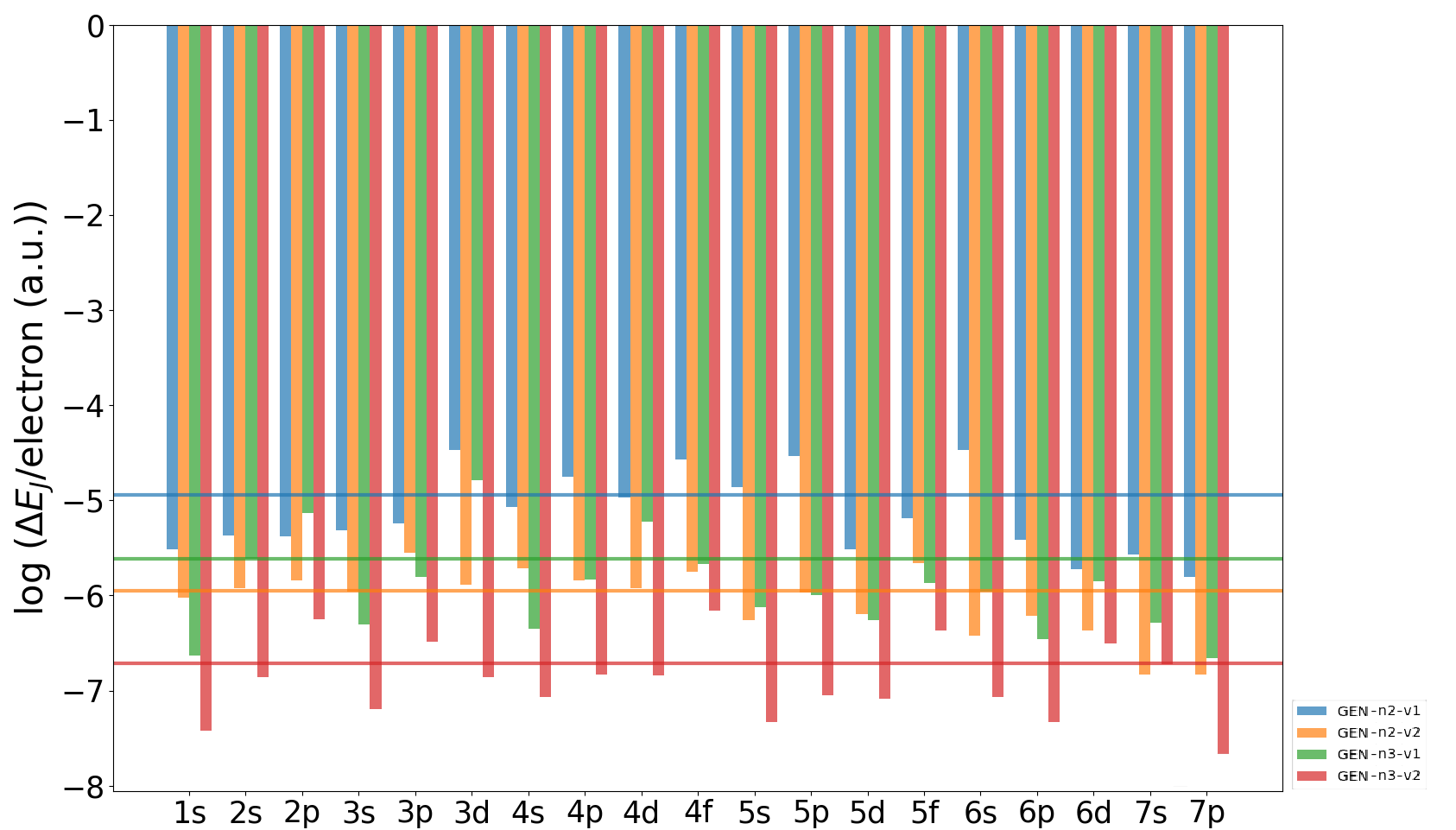}
    \caption{Mean value of $\Delta E_J/{\it per \ electron}$ in a.u. for each group of the molecular data base. The data are obtained using different ABSs, namely \texttt{GEN-n2-v1} (blue), \texttt{GEN-n2-v2} (yellow), \texttt{GEN-n3-v1} (green) and \texttt{GEN-n3-v2} (red). The horizontal lines with the same colors represent the mean value of $\Delta E_J/{\it per \ electron}$ for all molecular systems. The data are shown in logarithmic scale.}
    \label{histCoulEl}
\end{figure}

For the sake of completeness, the results are further summarised in Figures \ref{histCoul} and \ref{histCoulEl}
where we present the comparison of different auxiliary fitting basis sets visualised for each group of the molecular data set in the form of the mean value of $\Delta E_J$ and the mean value of $\Delta E_J/{\it per \ electron}$, respectively. The data are shown on a logarithmic scale.
\clearpage

\begin{figure}[H]
    \centering
    \includegraphics[width=1\textwidth]{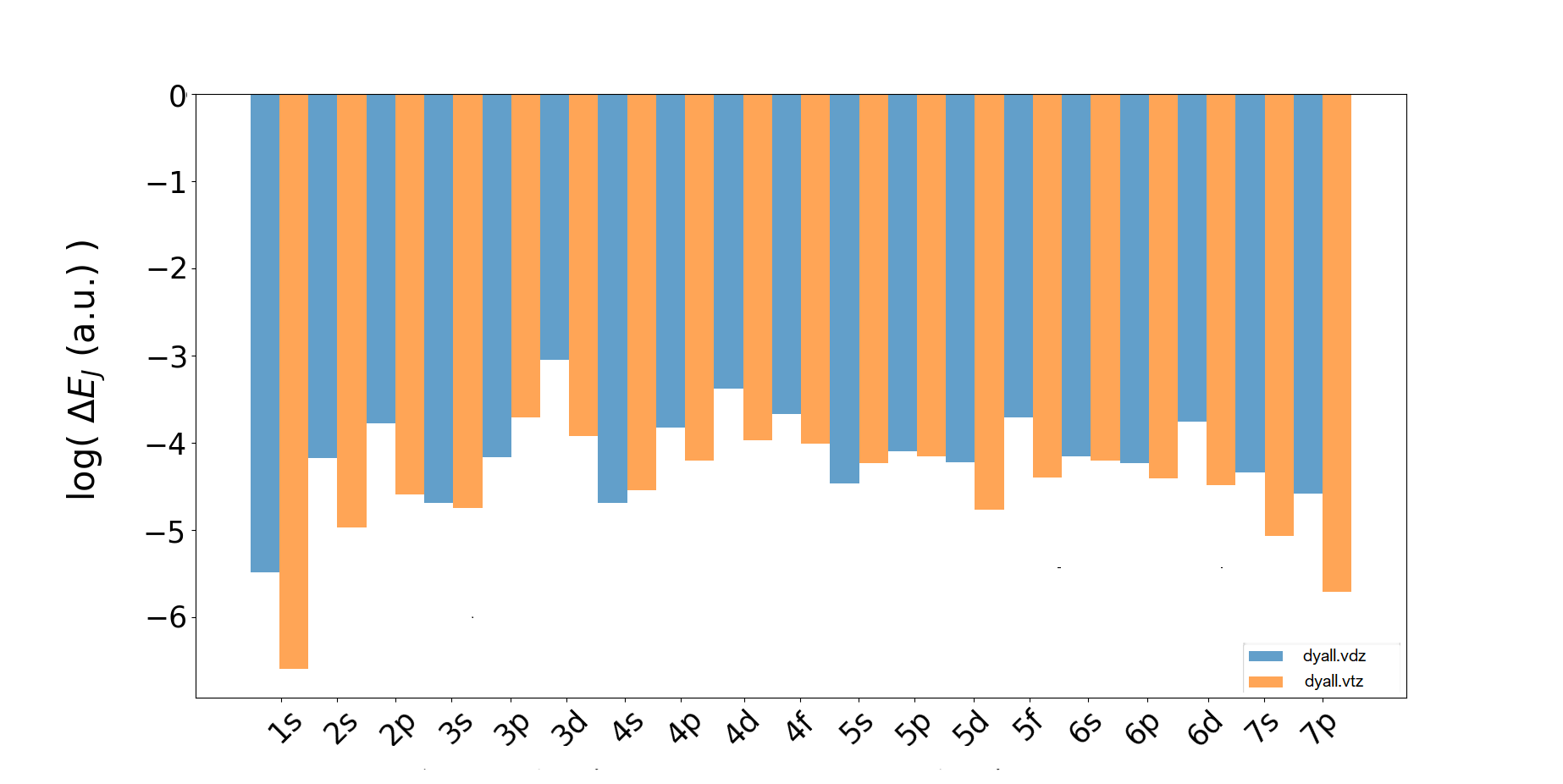}
    \caption{Comparison between the logarithms of the mean errors on $\Delta E_J$i n a.u. for the \texttt{GEN-n3-v1} fitting set generated from the dyall.v2z (in blue) and dyall.v3z (in orange) basis sets.}
    \label{histon3v1VTZVDZ}
\end{figure}

As mentioned above, one of the potential advantages of the automatic ABS generation process is that it can adapt to the main basis set used for the generation. In Figure~\ref{histon3v1VTZVDZ}, we clearly show that despite its simplicity, our automatic generation algorithm is able to provide accurate ABS even when we use a higher angular momentum principal basis set.
The data show the comparison of the mean error of $\Delta E_J$ obtained with \textit{dyall.v2z} and \textit{dyall.v3z} in combination with the respective automatically generated \texttt{GEN-n3-v1} basis sets. It is interesting to note that when switching from \textit{dyall.v2z} to \textit{dyall.v3z}, the accuracy is even slightly improved due to the higher angular flexibility of the generated \texttt{GEN-n3-v1} basis set.

%

To further asses our results it is interesting to compare the automatically generated fitting sets 
( namely \texttt{GEN-n2-v1}, \texttt{GEN-n2-v2}, \texttt{GEN-n3-v1} and \texttt{GEN-n3-v2} ) with respect to the two auxiliary basis sets that we optimised for gold atom\cite{BelpB16_20}, named $B16$ and $B20$. Indeed, in Table~\ref{tablespecconst} we show the results of the density fitting calculations using the four different ABSs generated from the \textit{dyall.v2z} and the ones obtained using the two cited $B16$ and $B20$ sets. As a reference, the results obtained with the conventional (without density fitting) DKS calculation of the Au$_2$ molecule are reported.

For the equilibrium bond length of Au$_2$, the results show that the fitting method essentially agrees with the exact ones, regardless of the auxiliary basis used.
The largest discrepancy exists for the smallest basis (\texttt{GEN-n2-v1}), which underestimates the bond length by 0.004~\AA.
The agreement for other spectroscopic parameters is also excellent: the error for the harmonic frequency is 1~cm$^{-1}$ with the \texttt{GEN-n2-v1} and almost exact values for the bases \texttt{GEN-n2-v2}, \texttt{GEN-n3-v2} and \texttt{B20}. Similarly, the dissociation energy is almost exact
(error of 0.001 eV) when the \texttt{GEN-n3-v2} and \texttt{B20} basis is used.

\begin{table}[ht!]
        
        \begin{center}
		\caption{Spectroscopic constants in the Au dimer calculated at the DKS/BLYP using the dyall.v2z basis set in combination with
  several auxiliary density fitting basis set. The absolute error in the Coulomb are also reported. 
  }
        \label{tablespecconst}       
        \begin{tabular}{l|ccccccc}
\hline \hline                
                
	   		          &  B16      & B20        &       GEN-n2-v1   &  GEN-n2-v2            &  GEN-n3-v1             &  GEN-n3-v2     &  Exact  \\
\hline
                            &           &            &                 &                     &                      &              &         \\  
   R$_e$(\AA)               &  2.543    & 2.543      &      2.539      &   2.543             &    2.544             &  2.543      &   2.543      \\  
                            &    &   &                 &                     &                      &              &         \\  
   $\omega_e$ (cm$^{-1}$)   &  171.0    & 169.6      &      170.5      &    169.6            &    169.1             &  169.4       &   169.5       \\  
                            &    &     &                 &                     &                      &              &         \\  
   D$_e$ (eV)        &   3.162   &  3.157     &      3.221      &    3.146            &     3.167            &  3.155       &  3.156  \\  
                            &    &    &                 &                     &                      &              &         \\  
   $\Delta E_J$             &  0.000094 & 0.000006   &   0.000345      &   0.000055          &    0.000026          &  0.000002    &         \\  

        \end{tabular}
        \end{center}
\end{table}

An obvious disadvantage of automatically generated ABS is that they contain a larger number of functions than those found for specially optimised sets. This affects the computational efficiency, which is the main goal when the density fitting scheme is introduced.
In Table~\ref{tablespeedup} we show that, as expected, the automatically generated
ABS are generally larger than the optimised ones.

The largest ABS, \texttt{GEN-n3-v2}, achieves an error that is one third of the error achieved with \texttt{B20}, but at the cost of approximately doubling the number of fitting functions and consequently  increasing the computational effort. With \texttt{B20}, the increase in speed (i.e., Speed-up) compared to the calculation without density fitting is slightly less than 100, while with \texttt{GEN-n3-v2} it is slightly more than 50.
If, on the other hand, the similarly sized \texttt{B16} and \texttt{GEN-n2-v1} ABS are compared, it becomes clear that the error in the Coulomb energy of \texttt{GEN-n2-v1} is about an order of magnitude higher, although the acceleration is almost identical. 
The \texttt{GEN-n3-v2} basis set offers the highest accuracy but it is reasonable to argue that for those seeking a balance between cost and accuracy, \texttt{GEN-n3-v1} is probably a good compromise, at least for this system.
Although our ABSs have not yet reached the computational efficiency of \texttt{B20} (which benefits from less, well-optimised fitting functions), it is noteworthy that the generation of ABSs for the entire periodic table for a given atomic principal spinor takes less than one second.

\begin{table}[ht]
\caption{\label{tablespeedup}
Size of auxiliary fitting basis set,
CPU times (s), speed ratio and accuracy in Coulomb energy (mhartree) for gold dimer (Au$_2$). Speed-up is calculated with respect the calculation without density fitting algorithm.  DKS/BLYP using the Dyall.v2z basis set. The absolute error in the Coulomb and in the total energy are also reported. All runs done with Intel Xeon CPU E5-2683 v4 2.1.
}
\begin{center}
\begin{tabular}{ccccc}
\hline\hline
ABS     & Size & Time (s) & Speed-up & $\Delta E_J$ \\
\hline
B16     &  420  & 11   & 128        & 0.094        \\
B20     &  614  &14.8  & 95         & 0.006        \\
GEN-n2-v1 &  554  & 12.5 &  113     & 0.345        \\
GEN-n2-v2 &  944  & 19.4 &   73     & 0.055        \\
GEN-n3-v1 &  734  & 16.3 &   86     & 0.026        \\
GEN-n3-v2 &  1248 & 25.8 &   54     & 0.002        \\
\hline\hline
\end{tabular}
\end{center}
\end{table}

The numerical stability of the solution of the linear systems containing the matrix $\bf A$ (the Coulomb matrix of the two-electron integrals of the auxiliary functions, see Eq.\ref{eq:Adv} and Eq.\ref{eq:Azw}) is a subtle point to consider.
Although the matrix $\bf A$ is in principle positive definite, it may be ill-conditioned due to finite precision arithmetic. The condition number indicates how sensitive the solution of a linear system is to small changes in the input data or the matrix itself. A matrix with a lower condition number is better conditioned, i.e. the numerical solutions are probably more stable. Conversely, a high condition number indicates a poorly conditioned matrix where the solutions are very sensitive to perturbations, which can lead to numerical instability in the density fitting scheme. As mentioned in the literature, the automatically generated ABS can be particularly susceptible to this problem and therefore a numerical scheme can be introduced to avoid this problem (see K\"{o}ster et al.\cite{Koster:2025} and references therein for a detailed description).
In the case of the Au$_2$ molecule, we evaluate the condition numbers of the matrix $\bf A$ for all different
ABSs. Their values are very high in all cases: about $10^{20}$ and $10^{22}$ for $B16$ and $B20$ and, as expected, even higher, $10^{25}$, $10^{28}$, $10^{26}$ and $10^{30}$ for the automatically generated GEN-n2-v1, GEN-n2-v2, GEN-n3-v1 and GEN-n3-v2 respectively.
The higher values of the condition numbers are for the automatically generated ABS.
Interestingly, despite the high condition numbers, we did not observe any numerical noise during the SCF procedure for the entire set of 286 molecules.

As a more rigorous proof of the numerical stability, we tested our automatically generated ABSs in a real-time TDDKS simulation. We use the PyBERTHART code \cite{de2020pyberthart},
the Python API for the integral kernel of BERTHA.
The real-time propagation of the DKS equation is extremely sensitive to any source of numerical problems.
Indeed, at each step of the time evolution, the DKS matrix must be evaluated many times and the total number of time steps can be in the order of thousands.

\begin{figure}[H]
    \centering
    \includegraphics[width=1\textwidth]{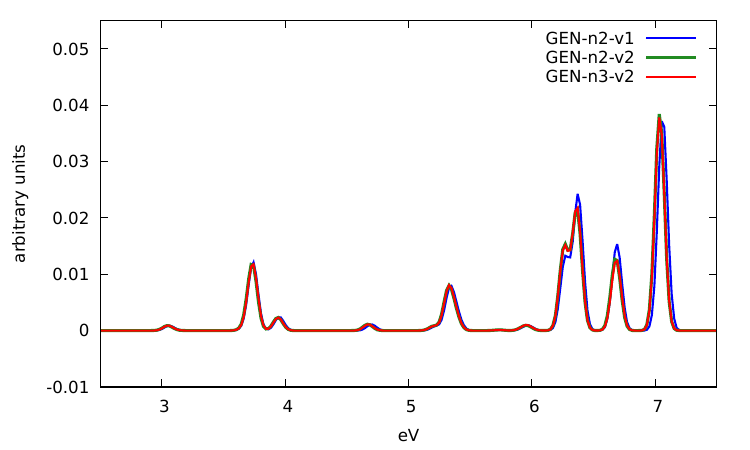}
    \caption{Absorption spectra of the PbCl$_2$ molecule using different automatic generated ABS. See text for details.}
    \label{berthart}
\end{figure}

In Figure \ref{berthart} we show the absorption spectrum of the PbCl$_2$ molecule using the dyall.2vz basis in combination with different automatically generated ABSs.
The spectrum was obtained by Fourier transforming the
 time-dependent dipole moment induced by a kick perturbation using a Gaussian dumping function with an exponent of $1\cdot 10^{-6}$. The simulation was performed with a time step of 0.1 a.u. for a total simulation time of 72.5 fs (i.e., 30000 time steps).
Regardless of the ABS used, we observed very stable propagation both in terms of induced dipole and total energy conservation.
The resulting absorption spectrum is very clean, without the numerical artefacts that typically occur with numerical instability of the time-evolution procedure.

\section{Conclusions}

The development of efficient algorithms based on density fitting is an active area of research as it can be crucial to reduce the computational cost of electronic structure methods in quantum chemistry. While this approach is well established in the non-relativistic regime, its full applicability in a four-component relativistic DKS framework remains a challenge, in particular due to the lack of well-optimised auxiliary basis sets (ABSs) for the heaviest elements of the periodic table.

In this work, we have developed a fully automatic and modular workflow for the generation of ABS in the framework of a relativistic 4-component code, starting from some information (exponents and angular distribution) obtained from the basis sets of the main atomic spinors. We have generated auxiliary basis sets of different sizes and accuracies and tested their performance in the DKS module of the BERTHA code using a large molecular data set (almost 300 molecules with elements from H to Og). With the largest auxiliary basis set, we can achieve an accuracy  in the
Coulomb energy even of less than a $\mu$Hartree. This is comparable to the accepted errors for the non-relativistic case. The choice of principal basis set has a significant impact on the size of the ABS, which consequently affects the computational cost and efficiency.
Remarkably, our approach maintains high accuracy over the entire periodic table with minimal deviations and remains stable even when large principal basis sets are used (e.g. the basis set dyall.vtz).


A key advantage of our workflow is its automation, which enables the rapid creation, testing and refinement of auxiliary basis sets. This makes it a valuable tool for future developments in the field of new density auxiliary basis sets generation.
While our results show high accuracy, our preliminary analysis shows that automatically generated ABS generally have higher condition numbers than fitting basis sets that have been specifically optimised according to energetic criteria.
Our future work should focus on improving the algorithm for the automatic generation of exponents. This could potentially lead to a further reduction in the size of such auxiliary basis sets and help avoiding potential numerical instabilities by using the reduction of the condition number as an additional criterion.
The algorithm proposed here is accurate for the DKS calculations with local (LDA) and semi-local (GGA) exchange correlation functionals. However, it is known that high accuracy for Fock exchange requires auxiliary basis sets with much higher angular flexibility.\cite{bagel:2013,Koster:2025}
Recently, K\"{o}ster et al.\cite{Koster:2025} developed an automatic generation of auxiliary basis sets of Hermite-Gaussian Type functions that are accurate for Hartree-Fock and DFT using hybrid functions for all elements up to Kr. The automatic workflow presented here forms the basis for future research into optimal auxiliary basis sets to extend four-component DKS to efficiently incorporate exact Fock exchange.

\begin{acknowledgement}
L. S. and N. A. acknowledge funding from “PaGUSci - Parallelization and GPU Porting of Scientific Codes”  CUP: C53C22000350006 within the Cascading Call issued by Fondazione  ICSC , Spoke 3 Astrophysics and Cosmos Observations. National Recovery and Resilience Plan (Piano Nazionale di Ripresa e Resilienza, PNRR) Project ID CN\_00000013 "Italian Research Center on High-Performance Computing, Big Data and Quantum Computing" funded by MUR Missione 4 Componente 2 Investimento 1.4: Potenziamento strutture di ricerca e creazione di "campioni nazionali di R\&S (M4C2-19 )" - Next Generation EU (NGEU).
E. R. acknowledges funding from the European Research Council (ERC) under the European Union’s Horizon Europe Research and Innovation Programme (Grant No. ERC-StG-2021-101040197—QED-Spin).
L. B. acknowledges  financial support from ICSC-Centro Nazionale di Ricerca in High Performance Computing, Big Data and Quantum Computing, founded by European Union-Next-Generation-UE-PNRR, Missione 4 Componente 2 Investimento 1.4. CUP: B93C22000620006. 
\end{acknowledgement}

\begin{suppinfo}

\end{suppinfo}

\bibliography{shortj,biblio}

\end{document}